\documentclass[%
]{mpi2015-cscpreprint}

\usepackage[american]{babel}

\usepackage{makecell}
\usepackage{subcaption}
\usepackage{float} 
\usepackage{amssymb,amsfonts,amsmath}
\usepackage{graphicx,color,changebar}
\usepackage{graphicx}    
\usepackage{tikz}
\usepackage{pgfplots}
\newlength\fheight
\newlength\fwidth

\usepackage{algorithmic}
\usepackage[linesnumbered,ruled,vlined]{algorithm2e}
\SetKwInput{KwInput}{Input}                
\SetKwInput{KwOutput}{Output}              

\usepackage{cleveref}

\newtheorem{rem}{Remark}
\newtheorem{defn}{Definition}[section]

\usetikzlibrary{calc,patterns,
	decorations.pathmorphing,
	decorations.markings}

\usepackage[hardware, software]{mymacros}

  \usepackage{pgfplots}
\newenvironment{customlegend}[1][]{%
	\begingroup
	\csname pgfplots@init@cleared@structures\endcsname
	\pgfplotsset{#1}%
}{  \csname pgfplots@createlegend\endcsname
	\endgroup
}

\def\addlegendimage{\csname pgfplots@addlegendimage\endcsname}

\usepackage{mymacros}

\begin{document}
  

\title{A Non-Intrusive Method to Inferring Linear Port-Hamiltonian Realizations using Time-Domain Data}
  
\author[$1$]{Karim Cherifi}
\affil[$1$]{Max Planck Institute for Dynamics of Complex Technical Systems, Sandtorstr.\ 1, 39106 Magdeburg, Germany.\authorcr
	\email{cherifi@mpi-magdeburg.mpg.de}, \orcid{0000-0000-0000-0000}}
	
\author[$2$]{Pawan Goyal}
\affil[$2$]{Max Planck Institute for Dynamics of Complex Technical Systems, Sandtorstr.\ 1, 39106 Magdeburg, Germany.\authorcr
	\email{goyalp@mpi-magdeburg.mpg.de}, \orcid{0000-0000-0000-0000}}
	
	\author[$3$]{Peter Benner}
	\affil [$3$]{Max Planck Institute for Dynamics of Complex Technical Systems, Sandtorstr.\ 1, 39106 Magdeburg, Germany.\authorcr
		\email{benner@mpi-magdeburg.mpg.de}, \orcid{0000-0000-0000-0000}}

\shorttitle{pH realization}
\shortauthor{K.Cherifi, P.Goyal, P.Benner}
  
\keywords{System identification, port-Hamiltonian systems, Loewner approach, transfer function, input-output data}

\msc{93A30, 93B30, 93B15, 93B20}
  
\abstract{%
Port-Hamiltonian systems have gained a lot of attention in recent years due to their inherent valuable properties in modeling and control. In this paper, we are interested in constructing linear port-Hamiltonian systems from time-domain input-output data. We discuss a non-intrusive methodology that is comprised of two main ingredients --- (a)  inferring frequency response data from time-domain data and (b) constructing an underlying port-Hamiltonian realization using the inferred frequency response data. 
We illustrate the proposed methodology by means of two numerical examples and also compare it with two other system identification methods to infer the frequency response from the input-output data.
	}
  
\maketitle


\section{Introduction}
In this paper, we focus on a non-intrusive way for the construction of a class of linear structured systems. Non-intrusive modeling  has received a lot of attention recently due to its data-driven nature, see, e.g., \cite{brunton2019data,yu2019non,guo2019data,PehW16}. There are primarily two fundamental ways to obtain data, leading to data-driven modeling. The first is to  experiment in a laboratory to obtain data. This approach is often desirable when very little knowledge is available about the process and parameters. In the second approach, one can simulate a process or model using proprietary software with desired parameters and conditions. Indeed, one can seek to obtain the underlying model in a matrix-vector form; however, it is a quite challenging task to extract the model, or sometimes even impossible due to intellectual property rights. Nevertheless, one can easily obtain simulated data using simulation software.  Anyhow, in both cases, the goal is to create a model that describes the data and incorporates (if available)  any additional knowledge such as conversation laws, and particular desired properties such as port-Hamiltonian structure.

In this work, our focus lies on inferring  \emph{linear time-invariant} (LTI) \emph{port-Hamiltonian} (pH) systems using  time-domain input-output data. PH systems are structured representations of dynamical systems
\cite{Sch06,JacZ12,OrtSMM01,Sch04,SchJ14}  that typically arise, e.g., from energy-based modeling via bond graphs \cite{GolSBM03,Bre08}. Constructing  compact and reduced-order models from  a complex large-scale pH system is a very active research area. However, all these methods require full knowledge about the model/process such as model parameters and discretization schemes. As discussed earlier, this may not be possible in several scenarios.  Therefore, we focus on  pH modeling using only input-output data. With this aim, the authors in \cite{Bea20,morBenGV20} proposed realization methods based on frequency-domain measurements. In \cite{Bea20}, the author presented a method that uses frequency response data to find a realization, not necessarily in a pH form; however, a linear matrix inequality (LMI) based approach was proposed to find an underlying pH realization. This extends the results presented in \cite{BeaMX15_ppt,BeaMV19}. On the other hand, the authors in \cite{morBenGV20} have extended the Loewner framework \cite{morMayA07} in order to directly obtain a pH model by the construction of the Loewner and shifted Loewner pencils in a particular way. However, the frequency response data of a system may not be readily available in some applications, see, e.g.,~\cite{DomHLT17}. In such applications, it might be easier to obtain time-domain experimental data or collect simulation data for a given input using proprietary software. 

One example, where a non-intrusive approach of modeling pH systems using time-domain data would be of great importance, is the modeling of gas transport networks. Although there exists a general hierarchy of submodels, see~\cite{DomHLT17}, there is no general model available. In particular, the compressor stations in gas networks do not have a first-principles model, but time-domain input-output data can be obtained. Then, in order to express the whole gas transport system as a single pH model using a  network of hierarchies of pH sub-models, a non-intrusive method is of high interest to generate a model for this component.
Another example, in this direction, is the modeling of a cable-driven parallel robot \cite{SchYSB18}. Such models are used to design physics-shaping controllers. Although there have been attempts to analytically build  pH models, the analysis in \cite{SchYSB18} shows the limitations and the complexity in order to derive analytically large and complex pH models.  Similar motivating examples can be found, e.g., in power electronics \cite{CupGBZJMD19} and continuous stirred tank reactors~\cite{HoaCJL11}.

Having noted various applications, our focus is to identify an underlying pH model of a process using  time-domain data. Since there exist tools to build pH models from frequency response data of a system \cite{morBenGV20}, our primary goal is to estimate the frequency response data using the time-domain data. For this, we first impose the linearity  assumption on the underlying dynamics. Under this assumption, there exist techniques that allow us to achieve our desired goal, see, e.g., \cite{PehGW17}. By combining the ideas discussed in \cite{morBenGV20} and \cite{PehGW17}, we propose a procedure to infer an underlying pH realization  using time-domain data obtained, e.g., using  proprietary software, or in an experimental set-up. Alternatively, one can construct an intermediate state-space realization using standard system identification techniques, e.g., the Multivariable Output-Error State Space (MOESP) method \cite{morVer94} and the Canonical Variate Algorithm (CVA) method \cite{morLar90}. Then, one can again use the results presented in \cite{morBenGV20} to find a pH realization.

The rest of the paper is structured as follows. In the following section, we introduce the pH framework and present the state of the art methods to infer pH realizations from data. Subsequently, in \Cref{sec:TD_loew}, we discuss a time-domain Loewner framework that allows us to infer an underlying pH-model using time-domain data. \Cref{sec:numerics} demonstrates the proposed procedure by means of two numerical examples and compares it with two different system identification methods: MOESP and CVA. In \Cref{sec:conclusion}, we conclude the paper with a short summary. 

\section{Port-Hamiltonian Systems and Previous Work}\label{sec:phSystems}
The pH framework is powerful as it inherently encodes underlying physical principles directly into the structure of the system model. An LTI pH system can be written in the following form \cite{BeaMV19}:
\begin{equation}\label{PHdef}
	\begin{aligned}
		\dot \bx(t) &=\left(\bJ-\bR\right)\bQ\bx(t) +(\bF-\bP)\bu(t), \\
		\by(t)&=(\bF+\bP)^\top \bQ\bx(t)  + (\bS+\bN) \bu(t) ,
	\end{aligned}
\end{equation}
where  $\bJ \in \Rnn$ is a skew-symmetric matrix, $\bR\in \Rnn$ is a positive semi-definite matrix, $\bF\pm \bP\in\Rnm$ are \emph{port} matrices, $\bS + \bN$ is the feed-through from the input to the output with $\bS=\bS^\top \geq 0 \in \mathbb \R^{m\times m} $ , $\bN=-\bN^\top \in \R^{m\times m}$ , and $\bQ\in \Rnn$ is a positive semi-definite matrix that is associated with the \emph{Hamiltonian} ${\mathcal H}(x)=\frac 12 \bx^\top\bQ\bx$. There is a close relationship between pH and passive systems, see, e.g.,~\cite{BeaMV19,ByrIW91}. To assure that an LTI system of the form \eqref{PHdef} is passive, the following condition is also required:
\begin{equation} \label{Kdef}
	\begin{bmatrix} \bR & \bP \\ \bP^\top & \bS \end{bmatrix}   \geq 0. \nonumber
\end{equation}

A model having the pH structure has many intrinsic  spectral properties~\cite{MehMW18} --- pH systems are robust under structured perturbations \cite{MehMS16,MehMS17}, pH systems are closed under power-conserving interconnection \cite{SchJ14}, and model reduction of large-scale pH systems via Galerkin projection yields low-order systems that inherit the pH structure, see, e.g.,~\cite{BeaG11,GugPBS12,PolS10}.

The main contributions, so far, in the direction of inferring pH realizations from data can be found in \cite{Bea20,morBenGV20,CheMH19}.
The methods in \cite{Bea20,morBenGV20} are based on frequency-domain data. The authors in \cite{morBenGV20} proposed a pH realization method from frequency-domain data based on the Loewner framework \cite{morMayA07}. However, the method requires data at the spectral zeros in the spectral directions which may not be easily available. Therefore, the authors in \cite{morBenGV20} proposed to infer first a state-space model using the standard Loewner approach and then to obtain samples at the spectral zeros in the spectral directions using the inferred state-space model. This approach is computationally efficient since an underlying pH system can be analytically determined. 
Recently, in \cite{Bea20}, a data-driven method to learning pH systems based on a solution of the passivity LMI~\cite{BeaMV19} is proposed. Moreover,
In \cite{CheMH19}, the authors proposed approaches to realize port-Hamiltonian systems using three different approaches. The goal is to learn a pH system in the form \eqref{PHdef} for given input-output time-domain data. In these methods, the idea is first to estimate frequency domain data as done in \cite{PehGW17}, followed by applying the Loewner approach to obtain a state-space model  of the form:
\begin{equation}\label{statesDef}
	\begin{aligned}
		\dot \bx(t) &=\bA\bx(t) +\bB\bu(t),\quad \bx(0) = 0, \\
		\by(t)&=\bC\bx(t)  + \bD \bu(t).
	\end{aligned}
\end{equation}
Then, given the system~\eqref{statesDef}, the authors seek an underlying pH structure and discuss approaches by means of, e.g., the solution of an LMI or the Lur\'e equations, or finding the nearest pH realization, belonging to the set of all admissible passive systems by solving the following optimization problem:
$$\underset{\bJ,\bR,\bQ,\bF,\bP,\bS}{\mathop{\inf }}\,\mathcal{G}(\bJ,\bR,\bQ,\bF,\bP,\bS)$$ subject to  ${{\bJ}^{\top}}=-\bJ,\bQ\ge 0\text{ and }\left[ \begin{matrix}
	\bR & \bP  \\
	{{\bP}^{\top}} & \bS  \\
\end{matrix} \right]\ge 0$, and where 
{\small 
	\begin{align*}
		&\mathcal{G}(\bJ,\bR,\bQ,\bF,\bP,\bS)=\left\| \bA-(\bJ-\bR)\bQ \right\|_{F}^{2}+\left\| \bB-(\bF-\bP) \right\|_{F}^{2} \\
		&\qquad\qquad\qquad\qquad~~+\left\| \bC-{{(\bF+\bP)}^{\top}}\bQ \right\|_{F}^{2}+\left\| \tfrac{\bD+{{\bD}^{\top}}}{2}-\bS \right\|_{F}^{2},
	\end{align*}
}
and $\bN$ is set to $\bN=\tfrac{\bD-{{\bD}^{\top}}}{2}$, see \cite{GilS18} for details.
This method allows for a general and versatile realization procedure. However, solving the above optimization problem is a challenging task. Towards this, the authors propose an algorithm to solve this optimization problem that is based on a fast projected gradient method (FGM) \cite{GilS18}. The method is in general faster than the standard projected gradient method for this type of problems where the objective function is non-convex. This also requires a restarting procedure to ensure that the algorithm converges. In addition, obtaining a suitable solution for this optimization problem requires choosing a good initial system.
The LMI based initialization procedure proposed in \cite{GilS18} works well when the initial system is close to being passive. However, the algorithm may get stuck in a local minimum.  Also, solving the LMI may be computationally expensive if the system is of large-scale. It is worth noting that only one representation of pH systems is considered. There may be a nearer pH system with another representation, e.g. in pH differential algebraic equation (DAE) form.  

In the following, we discuss a procedure that involves first inferring frequency response  data~\cite{PehGW17,CheMH19}. Then, we directly compute the underlying pH system using the method proposed in \cite{morBenGV20}. 

\section{Time Domain Loewner and Port-Hamiltonian Realization}\label{sec:TD_loew}
Our main objective is to realize a pH system from given time-domain input-output data, either obtained in an experimental set-up or by proprietary software. In essence, we first infer the frequency response data of the system using time-domain input-output data. 
This problem has been very well studied in the literature, see, e.g., \cite{morLjung87,IoAn12,Hokan13}, where frequency response data are typically inferred using the impulse response. Alternatively, one can compute a state-space realization using standard system identification methods \cite{morLjung87} and then use the resulting state-space realization to compute an equivalent pH realization. Among the most significant methods that can be used in this context are the MOESP method \cite{morVer94} and the CVA method \cite{morLar90}. 
%
However, to avoid computing the full state-space realization, we follow the method presented in \cite{PehGW17} to estimate the frequency data points at pre-defined interpolation points by designing an appropriate input. This is done by solving a least-squares problem. See \cite{PehGW17} for details. 

For this, let us assume that we excite the system using an input $\bu(t)$ and consider $K$ samples of $\bu(t)$ at the time $k T_s$, denoted as $\bu_k := \bu(kT_s)$, where $T_s$ is the sampling time and $k \in \{0,\ldots, K\}$. Then, using the discrete Fourier transform, we can write
\begin{equation*} 
	\bu_k=\sum_{i=0}^{K-1}{{{\bU}_{i}}\bq_{i}^{k}}, 
\end{equation*}
where 
${{\bU}_{i}}$ are the corresponding (discrete) Fourier coefficients, and ${{\bq}_{i}}={{e}^{\frac{2\pi \jmath}{K}i}}$ for $i \in \{0,\ldots,K-1\}$, $k \in \{0,\ldots,K-1\}$.
Moreover, under the linearity assumption, we can write the output sequence as follows:
\begin{equation} 
	{{\by}_{k}}=\sum\limits_{i\in {{\Gamma }_{r}}}{{{\bU}_{i}}{{\bH}_{k}}({{\bq}_{i}})\bq_{i}^{k}}, \label{output}
\end{equation}
where ${{\Gamma }_{r}=\{i_1,\ldots,i_r \}}$ are the indices of the $r$ non-zero Fourier coefficients ${\{{\bU}_{1},\ldots,{\bU}_{r}\}}$.
Equation~\eqref{output} gives us a direct relationship between the output $\by_k$ and the approximate frequency response data ${{\bH}_{k}}$ sampled at the frequency ${{\bq}_{i}}$.
To estimate the frequency response data of the system, one can solve a least-squares problem of the form:
\begin{equation}
	\widehat{\bH}=\underset{{{{\widehat{\bH}}}_{1}}^{'},\ldots,{{{\widehat{\bH}}}_{r}}^{'}\in \mathbb{C}}{\mathop{\arg \min }}\,\sum\limits_{k={{k}_{\min }}}^{K-1}
	\left(\by_k-\sum\limits_{l=1}^r{{\bU_{i_l}}{{{\widehat{\bH}}}_{l}}^{'}\bq_{{{i}_{l}}}^{k}}  \right)^2 \label{LSprob}
\end{equation}
with the solution \[\widehat{\bH}={{\left[{{\widehat{\bH}}_{1}},\ldots,{{\widehat{\bH}}_{r}}\right]}^{\top}}\] and $k_{\min}$ is chosen such that \eqref{LSprob} is over-determined and has a unique solution. This is generally true when the system reaches a steady state.
Equation \eqref{LSprob} can be rewritten as 
\begin{equation}
	\label{LSprob2}
	\underset{\widehat{\bH}\in {{\mathbb{C}}^{r}}}{\mathop{\arg \min }}\,\left\| \bF\widehat{\bH}-\bar{\by} \right\|_{2}^{2},
\end{equation} 	
where $\bF\in {{\mathbb{C}}^{(\bK-{{k}_{\min }})\times r}}$ is as follows :
\begin{equation}
	\bF=\left[ \begin{matrix}
		{{\bU}_{{{i}_{1}}}}\bq_{{{i}_{1}}}^{{{k}_{\min }}} & \ldots  & {{\bU}_{{{i}_{m}}}}q_{{{i}_{r}}}^{{{k}_{\min }}}  \\
		\vdots  & \ddots  & \vdots   \\
		{{\bU}_{{{i}_{1}}}}\bq_{{{i}_{1}}}^{K-1} & \ldots  & {{\bU}_{{{i}_{m}}}}q_{{{i}_{r}}}^{K-1}  \\
	\end{matrix} \right], \label{F}
\end{equation}
and $\bar{\by}={{\left[{{\by}_{{{k}_{\min }}}},\ldots,{{\by}_{K-1}}\right]}^{\top}}$. 
Once we estimate the frequency response data ${{\widehat{\bH}}_{1}},\ldots,{{\widehat{\bH}}_{r}}$ at the corresponding interpolation points ${{\bq}_{{{i}_{1}}}},\ldots,{{\bq}_{{{i}_{r}}}}$, the classical Loewner framework  can be employed \cite{morMayA07}.  For this, the first step is to select an even number $\widehat m \leq r$ of interpolation points and define $n := \tfrac{\widehat m}{2}$. Then,  the set of interpolation points is partitioned into left and right interpolation points: $ {{\bq}_{{{i}_{1}}}},\ldots,{{\bq}_{{{i}_{n}}}} $ and  ${{{\bq}_{{{i}_{n+1}}}},\ldots,{{\bq}_{{{i}_{\widehat m}}}}}$, respectively. Next, we can  construct the Loewner and shifted Loewner matrices as follows:
\begin{align*}
	{\widehat{\mathbb{L}}} &=\left[ \begin{matrix}
		\frac{{{{\widehat{\bH}}}_{1}}-{{{\widehat{\bH}}}_{n+1}}}{{{\bq}_{{{i}_{1}}}}-{{\bq}_{{{i}_{n+1}}}}} & \cdots  & \frac{{{{\widehat{\bH}}}_{1}}-{{{\widehat{\bH}}}_{2n}}}{{{\bq}_{{{i}_{1}}}}-{{\bq}_{{{i}_{2n}}}}}  \\
		\vdots  & \ddots  & \vdots   \\
		\frac{{{{\widehat{\bH}}}_{_{{{i}_{n}}}}}-{{{\widehat{\bH}}}_{{{i}_{2n}}}}}{{{\bq}_{{{i}_{n}}}}-{{\bq}_{{{i}_{n+1}}}}} & \cdots  & \frac{{{{\widehat{\bH}}}_{_{{{i}_{n}}}}}-{{{\widehat{\bH}}}_{{{i}_{2n}}}}}{{{\bq}_{{{i}_{n}}}}-{{\bq}_{{{i}_{2n}}}}}  \\
	\end{matrix} \right], \\
	{{\widehat{\mathbb{L}}}_{s}} & =\left[ \begin{matrix}
		\frac{{{\bq}_{{{i}_{1}}}}{{{\widehat{\bH}}}_{1}}-{{\bq}_{{{i}_{n+1}}}}{{{\widehat{\bH}}}_{n+1}}}{{{\bq}_{{{i}_{1}}}}-{{\bq}_{{{i}_{n+1}}}}} & \cdots  & \frac{{{\bq}_{{{i}_{1}}}}{{{\widehat{\bH}}}_{1}}-{{\bq}_{{{i}_{2n}}}}{{{\widehat{\bH}}}_{2n}}}{{{\bq}_{{{i}_{1}}}}-{{\bq}_{{{i}_{2n}}}}}  \\
		\vdots  & \ddots  & \vdots   \\
		\frac{{{\bq}_{{{i}_{n}}}}{{{\widehat{\bH}}}_{_{{{i}_{n}}}}}-{{\bq}_{{{i}_{n+1}}}}{{{\widehat{\bH}}}_{{{i}_{2n}}}}}{{{\bq}_{{{i}_{n}}}}-{{\bq}_{{{i}_{n+1}}}}} & \cdots  & \frac{{{\bq}_{{{i}_{n}}}}{{{\widehat{\bH}}}_{_{{{i}_{n}}}}}-{{\bq}_{{{i}_{2n}}}}{{{\widehat{\bH}}}_{{{i}_{2n}}}}}{{{\bq}_{{{i}_{n}}}}-{{\bq}_{{{i}_{2n}}}}}  \\
	\end{matrix} \right].
\end{align*} 
This allows us to infer a discrete-time model in generalized state-space form:
\begin{equation}\label{Discsys}
	\begin{aligned}
		\widehat{\bE}  \bx_{k+1} &=\widehat{\bA}\bx_{k} +\widehat{\bB}\bu_{k}, \\
		\by_{k}&=\widehat{\bC}\bx_{k} ,
	\end{aligned}
\end{equation}
where
$\widehat{\bE}= -{\widehat{\mathbb{L}}}$,
$\widehat{\bA}= -{{\widehat{\mathbb{L}}}_{s}}$,
$\widehat{\bB}= \left[ {{\widehat{\bH}}_{1}},\ldots,{{\widehat{\bH}}_{n}}\right] ^\top $,
$\widehat{\bC}= \left[{{\widehat{\bH}}_{n+1}},\ldots,{{\widehat{\bH}}_{2n}}\right]$ with the assumption that the pencil $(s\widehat{\mathbb{L}},{{\widehat{\mathbb{L}}}_{s}})$ is regular. If it is not regular, then there exists a lower-order model that interpolates the data which can be obtained by performing a compression step, see \cite{morMayA07} for a detailed discussion.

The choice of the input $\bu(t)$, the interpolation points and the number of samples $K$ should be chosen wisely so that the frequency response data can be estimated up to a satisfactory tolerance in the desired range.  Typically, the range of the possible  frequencies that can be chosen to estimate the data is:
\[\left[ \tfrac{2\pi }{K},\tfrac{2\pi (K-1)}{K} \right].\]
It shows that as the number of samples $K$ increases, the range of possible frequencies increases. Moreover, the input should also be carefully chosen to span the frequency range of interest and so that it has non-zero Fourier coefficients only for the frequencies corresponding to the pre-defined interpolation points. This also ensures that  the matrix $\bF$ in \eqref{F} has low dimensions because the input is sparse in the Fourier-domain. To this purpose, the input is generally chosen to be a sum of cosine and sine signals:
\begin{equation*} 
	{{\bu}_{k}}=\frac{1}{K}\sum\limits_{l=1}^{m}{(1+\jmath)}\left(\cos \left(\frac{2\pi {{i}_{l}}k}{K}\right)+\jmath \sin\left(\frac{2\pi {{i}_{l}}k}{K}\right)\right) ,
\end{equation*}
where $k \in \{0,\ldots, K-1\}$ and $i_l$ are the pre-defined indices of the non-zero Fourier coefficients. 
Note that the resulting system~\eqref{Discsys} is a discrete-time system since it is inferred using discrete sampling of input and output. However, a discrete-time system can be transformed into a continuous-time system based on the implicit Euler method where the discrete time frequency domain variable z is transformed to the continuous time frequency domain variable s with the relation:  $z=\tfrac{1}{1-sT_s}$ . The continuous time system is then computed as:
\begin{equation}\label{eq:d2c}
	\bE_c = \widehat{\bE}, \quad \bA_c = \tfrac{1}{T_s}(\widehat{\bA} - \widehat{\bE}) , \quad
	\bB_c = \tfrac{1}{T_s}\widehat{\bB}, \quad \bC_c = \widehat{\bC}. 
\end{equation}

Notice that in the computation of the discrete time realization $\widehat{\bD}=0$. However, in order to compute a pH realization we need a nonzero $\bD_c$ matrix. In practice, it is set to a small value $\bD_c=10^{-5}$ in order to regularize the system.

Once we have a realization of the system, we can construct a pH realization as discussed in \cite{morBenGV20}. In the following, we first define the spectral zeros and zero directions.
\begin{defn}
	Given a transfer function $\bH(s)$ of an order $n$ system, the pairs $(s_j,r_j), j \in \{1,\ldots,n\}$, are \emph{spectral zeros}  and \emph{zero directions} if 
	\begin{equation}\label{eq:spectral}
		\Phi ({{s}_{j}}){{r}_{j}}=0 ,
	\end{equation}
	where $\Phi (s):={{\bH}^{*}}(-s)+\bH(s)$ and $``*"$ denotes the conjugate transpose.
	For a general state-space representation, these spectral zeros and spectral directions can be computed by solving the generalized eigenvalue problem \cite{morSor05,morBenF06}:
	\begin{equation} \label{SpecZero}
		\begin{bmatrix} 0 & \bA_c & \bB_c \\ \bA_c^\top & 0 & \bC_c^\top \\ \bB_c^\top & \bC_c & \bD_c + \bD_c^\top  \end{bmatrix} \begin{bmatrix} p_j \\ q_j \\ r_j \end{bmatrix} = s_j 
		\begin{bmatrix} 0 & \bE_c & 0 \\ -\bE_c^\top & 0 & 0 \\ 0 & 0 & 0  \end{bmatrix} \begin{bmatrix} p_j \\ q_j \\ r_j \end{bmatrix}.
	\end{equation}
\end{defn}
Now, let us consider data as follows:
\begin{align*}
	\bH(\lambda_j)r_j = w_j, \quad j\in \{1,\ldots, n\},
\end{align*}
where $\lambda_j \in \C^+$, $j \in \{1,\ldots,n\}$, are the spectral zeros of $\bH(s)$ in the open right half-plane and $r_j$ are the corresponding zero directions. By making use of \eqref{eq:spectral}, we also have
\begin{align*}
	r_j^*\bH(-\lambda_j^*) = w_j^{*}, \quad j\in \{1,\ldots, n\}.
\end{align*}
Next, we define the right and left tangential interpolations set as $(\lambda_j,r_j,w_j)$ and $(-\lambda_j^*,r_j^*,w_j^*)$, respectively. 
Then, using these interpolation conditions, we can obtain the Loewner and shifted Loewner matrices as follows: 
\begin{align*}
	\mathbb{L} &=\left[ \begin{matrix}
		\frac{r_1^{*}w_1+w_1^{*}r_1}{{{\lambda }_{1}}+{\lambda^*_{1}}} & \cdots  & \frac{r_1^{*}w_n+w_1^{*}r_n}{{{\lambda }_{n}}+{\lambda^*_{1}}}  \\
		\vdots  & \ddots  & \vdots   \\
		\frac{r_n^{*}w_1+w_n^{*}r_1}{{{\lambda }_{1}}+{\lambda^*_{n}}} & \cdots  & \frac{r_n^{*}w_n+w_n^{*}r_n}{{{\lambda }_{n}}+{\lambda^*_{n}}}  \\
	\end{matrix} \right], \\
	\mathbb{L}_{s}&=\left[ \begin{matrix}
		\frac{{{\lambda }_{1}}r_1^{*}w_1+{\lambda^*_{1}}w_1^{*}r_1}{{{\lambda }_{1}}+{\lambda^*_{1}}} & \cdots  & \frac{{{\lambda }_{n}}r_1^{*}w_n+{\lambda^*_{1}}w_1^{*}r_n}{{{\lambda }_{n}}+{\lambda^*_{1}}}  \\
		\vdots  & \ddots  & \vdots   \\
		\frac{{{\lambda }_{1}}r_n^{*}w_1+{\lambda^*_{n}}w_n^{*}r_1}{{{\lambda }_{1}}+{\lambda^*_{n}}} & \cdots  & \frac{{{\lambda }_{n}}r_n^{*}w_n+{\lambda^*_{n}}w_n^{*}r_n}{{{\lambda }_{n}}+{\lambda^*_{n}}}  \\
	\end{matrix} \right].
\end{align*}
This allows us to construct a state-space realization that also matches the transfer function at infinity as follows:

\begin{equation}\label{eq:ph1}
	\begin{split}
		\bE_{pH} &= -\mathbb L, \quad \bA_{pH} = \mathbb L_s - \bR^*\bD\bR, \quad
		\bB_{pH} = -\bW^* - \bR^*\bD, \\ \bC_{pH}& = -\bW + \bD\bR, \quad \bD_{pH} = \bD. 
	\end{split}
\end{equation}

It is possible that the realization  \eqref{eq:ph1} is complex. However, there exists an orthogonal transformation, allowing us to write the realization \eqref{eq:ph1} as a real system, see  \cite{AntLI17}. Moreover, the system \eqref{eq:ph1} can be transformed into pH form, as shown in \eqref{PHdef}, by a similarity transformation and satisfies all the necessary properties for a pH system. For a more detailed discussion, we refer to \cite{morBenGV20}. The whole procedure is summarized in algorithm \ref{alg:Alg1} .

\begin{algorithm} 
	\DontPrintSemicolon
	
	\KwInput{$T_s,\bar{\bu},\bar{\by}$.}
	\KwOutput{$(\bE_{pH},\bA_{pH},\bB_{pH},\bC_{pH},\bD_{pH})$.}
	
	Infer frequency domain data $\widehat{\bH}$ based on \eqref{LSprob2} using the time domain data $(\bar{\bu},\bar{\by})$.\\
	
	Using $\widehat{\bH}$, compute the discrete time system $(\widehat\bE,\widehat\bA,\widehat\bB,\widehat\bC,\widehat\bD)$ as shown in \eqref{Discsys}. \\
	
	Construct the continuous time system $(\bE_{c},\bA_{c},\bB_{c},\bC_{c},\bD_{c})$ according to \eqref{eq:d2c}.  \\
	
	Compute the spectral zeros and zeros directions from the system $(\bE_{c},\bA_{c},\bB_{c},\bC_{c},\bD_{c})$. \\
	
	Use the spectral zeros as interpolation points to construct the pH realization $(\bE_{pH},\bA_{pH},\bB_{pH},\bC_{pH},\bD_{pH})$ as shown in \eqref{eq:ph1}.

	\caption{Time domain port-Hamiltonian realization}
	\label{alg:Alg1}
\end{algorithm}

\begin{rem}
	In practice, The direct feed-through term $\bD$ can be estimated by observing the behavior at high frequencies. Moreover, it can also be assessed using the step response of the system. The step response near the time $t=0$ corresponds to the direct feed-through. 
\end{rem}

\begin{rem}
	We note that we estimate the transfer function using the input-output data in a certain time interval. The accuracy of the estimation also depends on the sampling time. Precisely, if the sampling time is small, we can estimate the transfer function more accurately. 
\end{rem}

\section{Numerical experiments}\label{sec:numerics}
In this section, we discuss the efficiency of the proposed procedure to identify port-Hamiltonian realizations using time-domain input-output data by means of two numerical examples. Furthermore, we assume that the input and output of a system are measured with a sampling period ($T_s$), and we consider in total $K$ samples.  As discussed in \Cref{sec:TD_loew}, the input is chosen as a sum of sine and cosine functions of the form as follows:
\begin{equation} 
	{{\bu}_{k}}=\frac{1}{K}\sum\limits_{l=1}^{m}{(1+\jmath)}\left(\cos \left(\frac{2\pi {{i}_{l}}k}{K}\right)+\jmath \sin\left(\frac{2\pi {{i}_{l}}k}{K}\right)\right),
	\label{eq:input}
\end{equation}
where $k \in \{0,\ldots,K{-}1\}$, and $k_{\min}$ is set to $k_{\min} =\frac{1}{4}K$  to ensure that the system has reached a  steady state (approximately) after $k_{\min}$ \cite{PehGW17}. The quantity $\{i_{1},\ldots, i_{m}\}$ are the indices of the interpolation points that have to be chosen a priory for each example.
In order to compare the Loewner based method, we construct the intermediate discrete model using other identification methods, namely, MOESP and CVA. We make use of the \matlab~function \texttt{ssest} with an appropriate option setting to identify models using the MOESP and CVA methods respectively, and for this, we need a priori to specify the order of the model as an input. 
Once the discrete system is obtained, the same subsequent steps described in \Cref{sec:TD_loew} are followed.
All the numerical experiments are done on a \amd~Ryzen 7 PRO 4750U processor~CPU@1.7GHz, up to 8MB cache, 16 GB RAM, Ubuntu 20.04 LTS, \matlab Version 9.8.0.1323502(R2020a) 64-bit(glnxa64). 

\subsection{RLC circuit}
First, we consider an RLC ladder circuit with 100 resistors, capacitors and inductors \cite{morGugA03}.  To identify the system, we preset the number of input-output samples to be collected equal to $K=10~000$  with a sampling period $T_s = 10^{-2}$. Then, we select $m = 100$ interpolation points from  the set of frequencies ${{\bq}_{i}}={{e}^{\frac{2\pi \sqrt{-1}}{K}i}}$ with $i \in \{0,\ldots,K-1\}$. The $m$ points are selected as follows: $\frac{m}{2}$ points are chosen such that their indices $\{i_{1},\ldots, i_{m/2}\}$ are logarithmically equidistant in the range $\{0,\ldots,K-1\}$. This results in the choice of $\frac{m}{2}$  interpolation points $\{q_{i_1},\ldots, q_{i_{m/2}}\} \in \{q_0,\ldots, q_{K-1}\}$. The other $\frac{m}{2}$ interpolation points are selected such that the set is closed under complex conjugation. The input is constructed from the indices of these interpolation points $\{i_{1},\ldots, i_{m}\}$ as shown in \eqref{eq:input}. The output response of the system to the chosen input is measured and the input-output data is collected. 

The selected interpolation points are then used to compute a discrete-time realization using the Loewner framework, as explained in \Cref{sec:TD_loew}. To compare the proposed methodology, we use MOSESP and CVA methods to identify discrete-time systems using the same input-output data. Once the discrete-time system is obtained, we determine a continuous-time system based on the backward Euler method. Assuming that the continuous-time system is minimal and passive, we can determine the pH realization as described in \Cref{sec:TD_loew}. This procedure is repeated using different state-space realization orders.

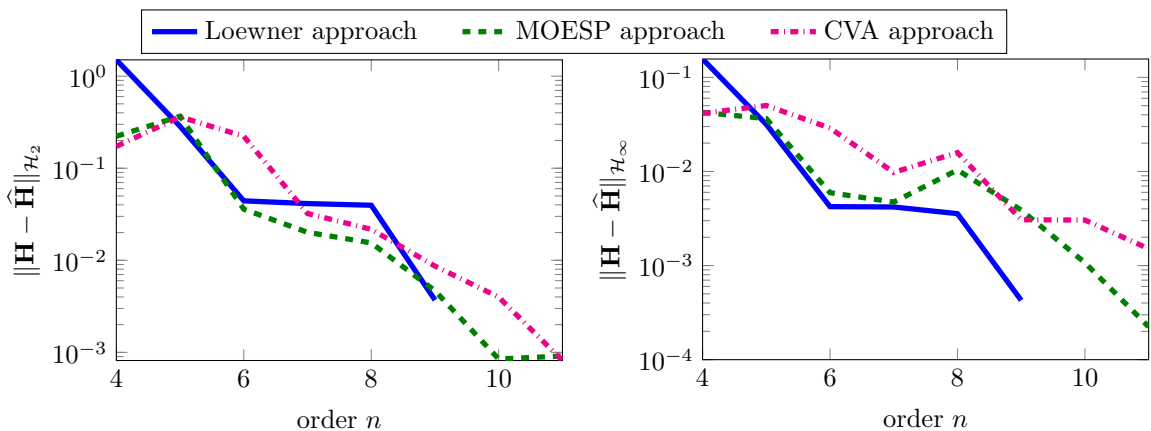
\begin{figure}[tb]
	\centering
	\begin{tikzpicture}
		\begin{customlegend}[legend columns=3, legend style={/tikz/every even column/.append style={column sep=0.5cm}} , legend entries={Loewner approach, MOESP approach, CVA approach}, ]
			\addlegendimage{color=blue, line width = 2pt}
			\addlegendimage{color=green!50!black,dashed, line width = 2pt}
			\addlegendimage{color=magenta,dashdotted, line width = 2pt}
		\end{customlegend}
	\end{tikzpicture}
	\begin{subfigure}{.5\textwidth}
		\centering
		\setlength\fheight{4.0cm}
		\setlength\fwidth{0.8\textwidth}
%
%
\begin{tikzpicture}

\begin{axis}[%
width=0.951\fwidth,
height=\fheight,
at={(0\fwidth,0\fheight)},
scale only axis,
xmin=4,
xmax=11,
ymode=log,
ymin=0.000815769431915733,
ymax=1.50276014213195,
yminorticks=true,
xlabel = {order $n$},
ylabel = {$\|\bH - \widehat{\bH}\|_{\cH_2}$},
axis background/.style={fill=white},
legend style={font=\tiny,at={(0.9,0.97)},legend cell align=left, align=left, draw=white!15!black}
]
\addplot [color=blue, line width = 2pt]
  table[row sep=crcr]{%
4	1.50276014213195\\
5	0.284977347926801\\
6	0.0442213916706224\\
7	0.041307229036043\\
8	0.0396219646112829\\
9	0.00370183534191791\\
};

\addplot [color=green!50!black, dashed, line width = 2pt]
  table[row sep=crcr]{%
4	0.222335896219975\\
5	0.366660379757141\\
6	0.0359177229242282\\
7	0.0201314433891319\\
8	0.0154646656886231\\
9	0.00466871042797187\\
10	0.000846093370687734\\
11	0.000913507629469522\\
};

\addplot [color=magenta, dashdotted, line width = 2pt]
  table[row sep=crcr]{%
4	0.172542656449038\\
5	0.361362315576819\\
6	0.222276508037515\\
7	0.0322179384348256\\
8	0.0216985450581676\\
9	0.00865077684768153\\
10	0.00398285370260077\\
11	0.000815769431915733\\
};

\end{axis}
\end{tikzpicture}%
	\end{subfigure}%
	\begin{subfigure}{.5\textwidth}
		\centering
		\setlength\fheight{4.0cm}
		\setlength\fwidth{0.8\textwidth}
%
%
\begin{tikzpicture}

\begin{axis}[%
width=0.951\fwidth,
height=\fheight,
at={(0\fwidth,0\fheight)},
scale only axis,
xmin=4,
xmax=11,
ymode=log,
ymin=0.0001,
ymax=0.156595685161994,
yminorticks=true,
xlabel = {order $n$},
ylabel = {$\|\bH- \widehat{\bH} \|_{\cH_{\infty}}$},
axis background/.style={fill=white},
legend style={font=\tiny,at={(0.9,0.97)},legend cell align=left, align=left, draw=white!15!black}
]
\addplot [color=blue,line width = 2pt]
  table[row sep=crcr]{%
4	0.156595685161994\\
5	0.0316468084766931\\
6	0.00422791826268783\\
7	0.00419274681526754\\
8	0.00356045784528796\\
9	0.000427178402849937\\
};

\addplot [color=green!50!black, dashed,line width = 2pt]
  table[row sep=crcr]{%
4	0.0424028213699904\\
5	0.0364368355412421\\
6	0.00594703719131593\\
7	0.00472691906351454\\
8	0.0103671062402668\\
9	0.00388703882279597\\
10	0.00107246246693649\\
11	0.000223674763861315\\
};

\addplot [color=magenta, dashdotted,line width = 2pt]
  table[row sep=crcr]{%
4	0.0414864834739959\\
5	0.0502073282198015\\
6	0.0289020345655757\\
7	0.00977540654301379\\
8	0.0158716736370934\\
9	0.00306839338364284\\
10	0.00304430398573508\\
11	0.00150932894280231\\
};

\end{axis}
\end{tikzpicture}%
	\end{subfigure}
	\caption{RLC circuit: A comparison of the $\cH_2$ and $\cH_{\infty}$-norms of the error between the original and various identified models for different orders.}
	\label{fig:RLCerror}
\end{figure}

To compare the quality of these methods, we compute the $\cH_2$ and $\cH_{\infty}$ errors between the original and identified models, which are reported in \Cref{fig:RLCerror}. It shows that the $\cH_2$ and $\cH_{\infty}$ errors with respect to the order of the obtained realization using various approaches. The figure suggests that there is no superior performance of an approach over the others, and the quality of the identified models depends on the chosen order as well as on the approach. 

However, we would like to emphasize that a suitable choice of order of the realization needs to be given for MOESP and CVA as an input, which is not known in advance. On the other hand, using the Loewner approach, we can determine a suitable order based on the decay of the singular values of the Loewner pencil.  It is a substantial advantage of the Loewner approach compared to the other methods. 

\begin{rem}
	Notice that the $\cH_2$ and $\cH_{\infty}$-norms for the Loewner approach in \Cref{fig:RLCerror} does not go beyond order $n = 9$. This is due to the truncation step that is implemented in the Loewner approach. Any order that is higher than $9$ will be truncated as the corresponding singular values are very small.
\end{rem}

\begin{figure}[tb]
	\begin{tikzpicture}
		\begin{customlegend}[legend columns=4, legend style={/tikz/every even column/.append style={column sep=0.5cm}} , legend entries={Original system, Loewner approach, MOESP approach, CVA approach}, ]
			\addlegendimage{color=red, only marks, mark=asterisk, mark options={solid, red}, line width = 2pt}
			\addlegendimage{color=blue, line width = 2pt}
			\addlegendimage{color=green!50!black,dashed, line width = 2pt}
			\addlegendimage{color=magenta,dashdotted, line width = 2pt}
		\end{customlegend}
	\end{tikzpicture}
	\centering
	\setlength\fheight{4.0cm}
	\setlength\fwidth{0.9\textwidth}
	\input{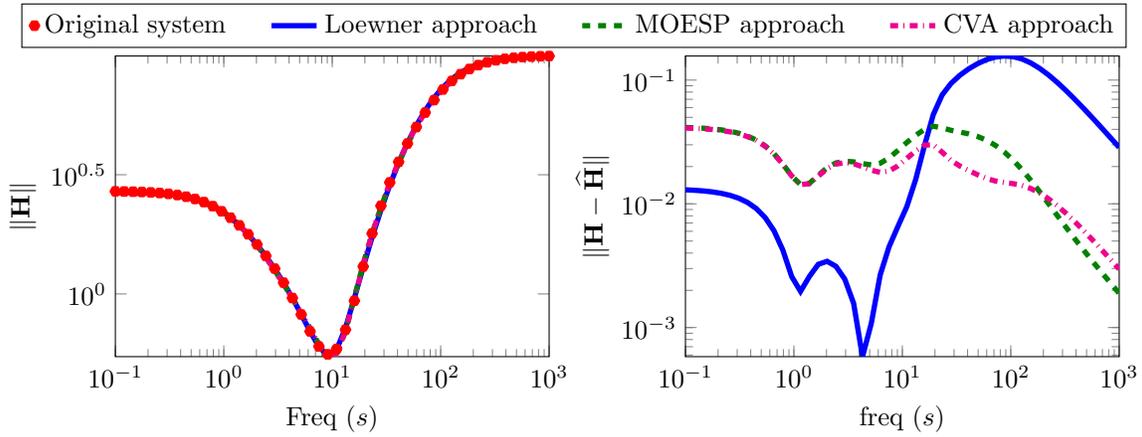}
	\caption{RLC circuit: A comparison for the Bode plot of the original and the realized pH systems.}
	\label{fig:NetBode}
\end{figure}

\begin{figure}[tb]
	\centering
	\begin{tikzpicture}
		\begin{customlegend}[legend columns=4, legend style={/tikz/every even column/.append style={column sep=0.5cm}} , legend entries={Original system, Loewner approach, MOESP approach, CVA approach}, ]
			\addlegendimage{color=red, only marks, mark=asterisk, mark options={solid, red}, line width = 2pt}
			\addlegendimage{color=blue, line width = 2pt}
			\addlegendimage{color=green!50!black,dashed, line width = 2pt}
			\addlegendimage{color=magenta,dashdotted, line width = 2pt}
		\end{customlegend}
	\end{tikzpicture}
	\setlength\fheight{4.0cm}
	\setlength\fwidth{0.9\textwidth}
	\input{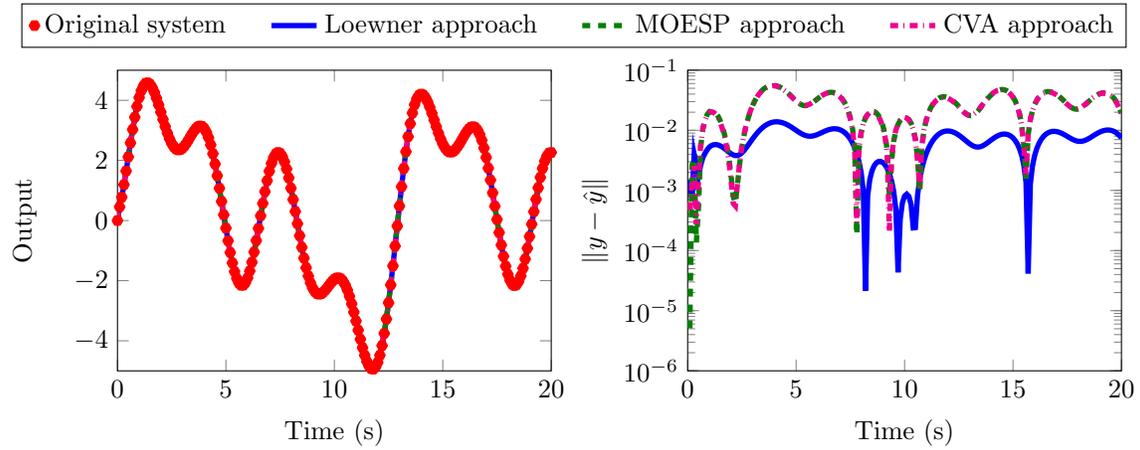}
	\caption{RLC circuit: A comparison for the time response of the original and the realized pH systems. }
	\label{fig:NetTime}
\end{figure}

Furthermore, to compare the transfer functions of the original and identified models, we construct models using various approaches by setting the order $n = 4$. The Bode plots of the original and the realized pH systems are shown in \Cref{fig:NetBode}. The figure shows that all three methods can be used to identity the system using input-output data. Moreover, the quality of all these approaches is comparable. Furthermore, we compare the transient response of the original and realized models using the input:
\begin{equation} \label{inputRLC}
	u(t)=\sin(t)+\sin(2 t)+\sin(0.5 t).
\end{equation} 
The resulting response is shown in \Cref{fig:NetTime}. It shows that all the realized models follow the original system very well.


In addition, we make a note of the computational cost of the approaches that is listed in \Cref{tab: RLC_CPU}.  In our vanilla implementation, we observe that the Loewner method was much faster than the other two methods. A potential reason for the Loewner approach to be faster could be that we solve a customized optimization problem to identify the transfer function at the specific frequencies that are present in the input. On the other hand, MOESP and CVA methods are more general and do not benefit from the specific choice of frequencies. Hence, the Loewner method can be of particular interest for a large data set, when the input contains a few frequency components.

\begin{table}[tb]
	\centering
	\begin{tabular}{| c| c | c | c |} \hline
		Method & Loewner & MOESP & CVA \\ \hline
		CPU Time & 0.15s & 4.87s  & 3.36s \\ \hline
	\end{tabular}
	\caption{RLC circuit: A comparison of the computational time to identify a model using the given data. }
	\label{tab: RLC_CPU}
\end{table}

\subsection*{Measurement noise}
Next, we study the performance of the approaches when input-output data is corrupted with measurement noise.  For this, we corrupt the output by adding Gaussian white noise of a certain level. Note that the norm of noise at a certain time is added relative to the magnitude of the output at that instance. To measure the performance of the approaches, we compare the  $\cH_2$ and $\cH_\infty$ norms of the error system between the original and identified realizations by considering different levels of noise. We identify the realization of order $n = 4$ by the Loewner, MOESP, and CVA approaches, and we change the level of the noise by considering a Gaussian white noise with different standard deviations, i.e., $\sigma \in \{10^{-2},10^{-3},10^{-4},10^{-5},10^{-6}\}$. The $\cH_2$ and $\cH_\infty$-norms of the error systems are compared in  \Cref{Table:H2RLC,Table:HinfRLC}. This empirical study shows that the MOESP and CVA approaches are comparable, and the Loewner approach yields slightly poorer results. However, overall, these approaches work very well under noise, and this experiment shows their robustness behavior with respect to the noise in measurement data. 

\begin{center} 
	\begin{table}
		\begin{tabular}{| c | c | c | c | c | c | c |}
			\hline
			& \multicolumn{6}{c}{ Standard deviation of noise ($\sigma$) } \vline \\ \hline
			Method & 0 & $10^{-6}$ & $10^{-5}$ & $10^{-4}$ & $10^{-3}$ & $10^{-2}$ \\ \hline
			Loewner  &	$2.13\cdot10^{-2}$ & $2.13\cdot10^{-2}$ & $2.13\cdot10^{-2}$ & $2.13\cdot10^{-2}$ & $2.25\cdot10^{-2}$ & $1.82\cdot10^{-2}$\\ \hline
			MOESP  & $5.50\cdot10^{-3}$ & $5.60\cdot10^{-3}$ & $5.70\cdot10^{-3}$ & $4.30\cdot10^{-3}$ & $1.13\cdot10^{-2}$ & $2.89\cdot10^{-2}$\\ \hline
			CVA & $7.10\cdot10^{-3}$ & $7.10\cdot10^{-3}$ & $7.50\cdot10^{-3}$ & $7.10\cdot10^{-3}$ & $8.00\cdot10^{-3}$ & $4.23\cdot10^{-2}$ \\ \hline	
		\end{tabular}
		\caption{RLC circuit: A comparison of the $\cH_2$-norm of the error between the original and realized systems under different levels of noise in the measurement data.}
		\label{Table:H2RLC}
	\end{table}
\end{center}

\begin{center} 
	\begin{table}
		\begin{tabular}{| c | c | c | c | c | c | c |}
			\hline
			& \multicolumn{6}{c}{ Standard deviation of noise ($\sigma$) } \vline \\ \hline
			Method & 0 & $10^{-6}$ & $10^{-5}$ & $10^{-4}$ & $10^{-3}$ & $10^{-2}$ \\ \hline
			Loewner & $1.60\cdot10^{-2}$ & $1.60\cdot10^{-2}$ & $1.60\cdot10^{-2}$ & $1.60\cdot10^{-2}$ & $1.72\cdot10^{-2}$ & $2.69\cdot10^{-2}$\\ \hline
			MOESP  & $5.70\cdot10^{-3}$ & $5.70\cdot10^{-3}$ & $5.70\cdot10^{-3}$ & $5.90\cdot10^{-3}$ & $7.71\cdot10^{-2}$ & $3.53\cdot10^{-2}$\\ \hline
			CVA  & $6.10\cdot10^{-3}$ & $6.10\cdot10^{-3}$ & $5.50\cdot10^{-3}$ & $6.10\cdot10^{-3}$ & $5.80\cdot10^{-3}$ & $4.40\cdot10^{-2}$\\ \hline		
		\end{tabular}
		\caption{RLC circuit: A comparison of the $\cH_\infty$-norm of the error between the original and realized systems under different levels of noise in the measurement data.}
		\label{Table:HinfRLC}
	\end{table}
\end{center}

\subsection{Spiral Inductor PEEC Model}
As the next example, we consider modeling a proximity sensor constituted of a spiral inductor and a plane of copper on top of the spiral. An equivalent circuit is obtained using the partial element equivalent circuit (PEEC) technique and that is used to model the system as described in \cite{morLiK05}. The model is available as part of the morwiki benchmark collection \cite{morWiki}. 

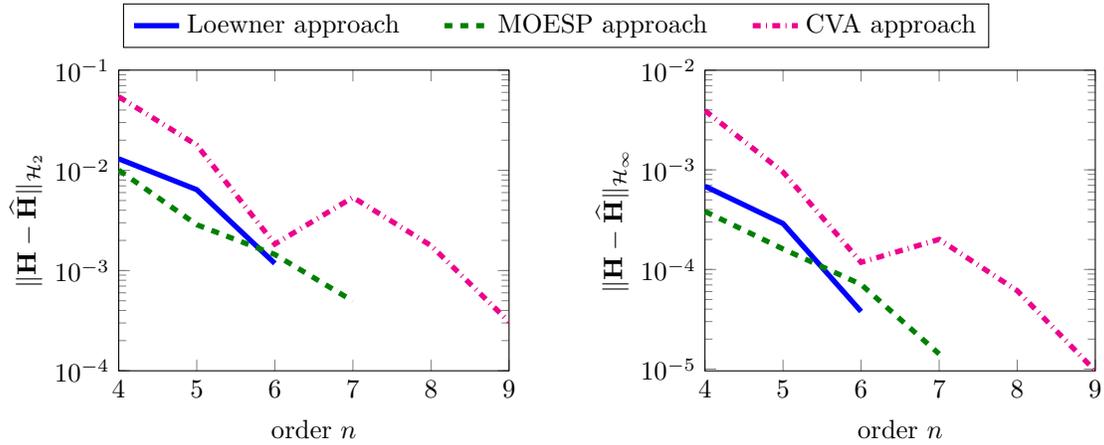
\begin{figure}
	\centering
	\begin{tikzpicture}
		\begin{customlegend}[legend columns=3, legend style={/tikz/every even column/.append style={column sep=0.5cm}} , legend entries={Loewner approach, MOESP approach, CVA approach}, ]
			\addlegendimage{color=blue, line width = 2pt}
			\addlegendimage{color=green!50!black,dashed, line width = 2pt}
			\addlegendimage{color=magenta,dashdotted, line width = 2pt}
		\end{customlegend}
	\end{tikzpicture}
	\begin{subfigure}{.5\textwidth}
		\centering
		\setlength\fheight{4.0cm}
		\setlength\fwidth{0.7\textwidth}
%
%
\begin{tikzpicture}

\begin{axis}[%
width=0.951\fwidth,
height=\fheight,
at={(0\fwidth,0\fheight)},
scale only axis,
xmin=4,
xmax=9,
ymode=log,
ymin=0.0001,
ymax=0.1,
yminorticks=true,
xlabel = {order $n$},
ylabel = {$\|\bH - \widehat{\bH} \|_{\cH_2}$},
axis background/.style={fill=white},
legend style={font=\tiny,at={(0.9,0.97)},legend cell align=left, align=left, draw=white!15!black}
]
\addplot [color=blue, line width = 2pt]
  table[row sep=crcr]{%
4	0.01303657221013\\
5	0.00641263227558184\\
6	0.00117847968675701\\
};

\addplot [color=green!50!black, dashed, line width = 2pt]
  table[row sep=crcr]{%
4	0.010062893575427\\
5	0.00286816580553397\\
6	0.0014453540018915\\
7	0.000496657348392036\\
};

\addplot [color=magenta, dashdotted, line width = 2pt]
  table[row sep=crcr]{%
4	0.0543570785185597\\
5	0.0178523638338205\\
6	0.00183853848952456\\
7	0.00537750727916182\\
8	0.00177521178455739\\
9	0.00030831579730021\\
};

\end{axis}
\end{tikzpicture}%
	\end{subfigure}%
	\begin{subfigure}{.5\textwidth}
		\centering
		\setlength\fheight{4.0cm}
		\setlength\fwidth{0.7\textwidth}
%
%
\begin{tikzpicture}

\begin{axis}[%
width=0.951\fwidth,
height=\fheight,
at={(0\fwidth,0\fheight)},
scale only axis,
xmin=4,
xmax=9,
ymode=log,
ymin=9.60075739265643e-06,
ymax=0.01,
yminorticks=true,
xlabel = {order $n$},
ylabel = {$\|\bH - \widehat{\bH} \|_{\cH_{\infty}}$},
axis background/.style={fill=white},
legend style={font=\tiny,at={(0.9,0.97)},legend cell align=left, align=left, draw=white!15!black}
]
\addplot [color=blue, line width = 2pt]
  table[row sep=crcr]{%
4	0.000685546299046568\\
5	0.000289284207420285\\
6	3.7891782026028e-05\\
};

\addplot [color=green!50!black, dashed, line width = 2pt]
  table[row sep=crcr]{%
4	0.000382697812200604\\
5	0.000160692148773785\\
6	7.08234301273784e-05\\
7	1.42048801165439e-05\\
};

\addplot [color=magenta, dashdotted, line width = 2pt]
  table[row sep=crcr]{%
4	0.0039369887320043\\
5	0.000950990762054891\\
6	0.000118006953077972\\
7	0.000200383290179501\\
8	6.12398207376682e-05\\
9	9.60075739265643e-06\\
};

\end{axis}
\end{tikzpicture}%
	\end{subfigure}%
	\caption{Spiral inductor: A comparison of $\cH_2$ and $\cH_{\infty}$ errors for different system's orders.}
	\label{fig:HSpiral}
\end{figure}

For this example, we also collected $10~000$ input-output samples as done in the previous example. Next, we identify the state-space realization of different orders by employing the Loewner, MOESP, and CVA approaches. To assess the performance of these approaches, we compare the $\cH_2$ and $\cH_\infty$-norms of the error between the original and identified realizations in \Cref{fig:HSpiral}.  We notice that the maximum order achieved using the Loewner, MOESP, and CVA is 6,7 and 8, respectively. Moreover,  like the previous example, we observe that there is no clear superior method among the three considered approaches, even for this example. 

\begin{figure}
	\begin{tikzpicture}
		\begin{customlegend}[legend columns=4, legend style={/tikz/every even column/.append style={column sep=0.5cm}} , legend entries={Original system, Loewner approach, MOESP approach, CVA approach}, ]
			\addlegendimage{color=red, only marks, mark=asterisk, mark options={solid, red}, line width = 2pt}
			\addlegendimage{color=blue, line width = 2pt}
			\addlegendimage{color=green!50!black,dashed, line width = 2pt}
			\addlegendimage{color=magenta,dashdotted, line width = 2pt}
		\end{customlegend}
	\end{tikzpicture}
	\centering
	\setlength\fheight{4.0cm}
	\setlength\fwidth{0.9\textwidth}
	\input{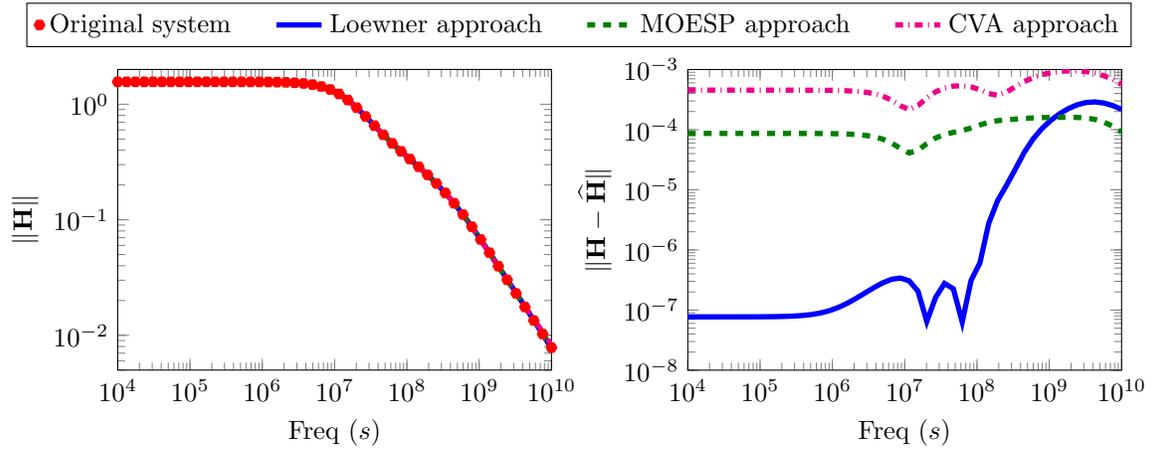}
	\caption{Spiral inductor: A comparison for the Bode plot of the original and the realized pH systems.}
	\label{fig:SpiralBode}
\end{figure}

\begin{figure}
	\begin{tikzpicture}
		\begin{customlegend}[legend columns=4, legend style={/tikz/every even column/.append style={column sep=0.5cm}} , legend entries={Original system, Loewner approach, MOESP approach, CVA approach}, ]
			\addlegendimage{color=red, only marks, mark=asterisk, mark options={solid, red}, line width = 2pt}
			\addlegendimage{color=blue, line width = 2pt}
			\addlegendimage{color=green!50!black,dashed, line width = 2pt}
			\addlegendimage{color=magenta,dashdotted, line width = 2pt}
		\end{customlegend}
	\end{tikzpicture}
	\centering
	\setlength\fheight{4.0cm}
	\setlength\fwidth{0.9\textwidth}
	\input{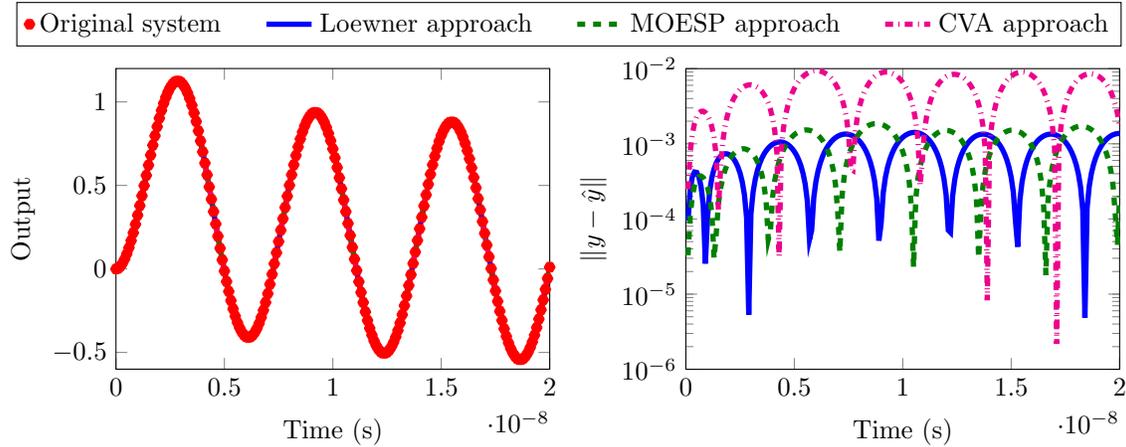}
	\caption{Spiral inductor: A comparison for the time response of the original and the realized pH systems.}
	\label{fig:SpiralTime}
\end{figure}

Next, for illustration, we plot the transfer functions of the identified realizations using the considered approaches for order $n=5$ in \Cref{fig:SpiralBode}.  The plot shows that all methods faithfully capture the behavior of the transfer function very well; however, in this case, the Loewner method seems to perform better. Moreover, we  compare the transient responses of the realized systems with the original system's one in \Cref{fig:SpiralTime} using the input \[\displaystyle{ u(t)=10(\sin(t)+\sin(10^9t)+\sin(0.25t))e^{-0.1t}}.\] 
Here also, we observe that all realized systems follow the transient behavior of the original systems, with the Loewner realization being slightly better compared to the other considered approaches. It is important to notice that using higher orders not achievable by the Loewner approach, slightly smaller errors can be obtained using MOESP and CVA methods.

Moreover, we again report the computational time required to identify the models using various approaches in \Cref{tab:spiral_CPU}, where we note that the Loewner approach needed the least computational time. 

\begin{table}[tb]
	\centering
	\begin{tabular}{| c| c | c | c |} \hline
		Method & Loewner & MOESP & CVA \\ \hline
		CPU Time & 0.14s & 6.07s  & 6.05s \\ \hline
	\end{tabular}
	\caption{Spiral inductor: A comparison of the computational time to identify a model using the given data. }
	\label{tab:spiral_CPU}
\end{table}

\subsection*{Measurement noise} In order to test the robustness of the methods to noise, we conduct the same experiment described in the first example. For this scenario, we fix the order to $n = 4$ and identify the realization by employing all three approaches under data being corrupted with Gaussian white noise. We corrupt the data, as described in the previous example. Next, in  \Cref{Table:H2Spiral,Table:HinfSpiral}, we compare the performance of the approaches by computing the $\cH_2$ and $\cH_\infty$-norms of the error systems. The tables indicate that all the methods are quite robust with respect to noise, and the Loewner approach provides sightly better models under noise as well. 

\begin{center} 
	\begin{table}
		\begin{tabular}{| c | c | c | c | c | c | c |}
			\hline
			& \multicolumn{6}{c}{ Standard deviation of noise ($\sigma$) } \vline \\ \hline
			Method & 0 & $10^{-6}$ & $10^{-5}$ & $10^{-4}$ & $10^{-3}$ & $10^{-2}$ \\ \hline	
			Loewner & $1.10\cdot10^{-2}$ & $1.10\cdot10^{-2}$ & $1.10\cdot10^{-2}$ & $1.10\cdot10^{-2}$ & $1.12\cdot10^{-2}$ & $1.50\cdot10^{-2}$ \\ \hline
			MOESP  & $5.11\cdot10^{-2}$ & $5.11\cdot10^{-2}$ & $5.11\cdot10^{-2}$ & $5.11\cdot10^{-2}$ & $5.07\cdot10^{-2}$ & $5.09\cdot10^{-2}$ \\ \hline
			CVA   & $4.41\cdot10^{-2}$ & $4.78\cdot10^{-2}$ & $4.89\cdot10^{-2}$ & $5.03\cdot10^{-2}$ & $5.07\cdot10^{-2}$ & $4.93\cdot10^{-2}$\\ \hline
		\end{tabular}
		\caption{Spiral inductor: A comparison of the $\cH_2$-norm of the error between the original and realized system using various approaches under different levels of noise in the measurement data.}
		\label{Table:H2Spiral}
	\end{table}
\end{center}

\begin{center} 
	\begin{table}
		\begin{tabular}{| c | c | c | c | c | c | c |}
			\hline
			& \multicolumn{6}{c}{ Standard deviation of noise ($\sigma$) } \vline \\ \hline
			Method & 0 & $10^{-6}$ & $10^{-5}$ & $10^{-4}$ & $10^{-3}$ & $10^{-2}$ \\ \hline
			Loewner & $6.92\cdot10^{-4}$ & $6.92\cdot10^{-4}$  & $6.92\cdot10^{-4}$  & $6.91\cdot10^{-4}$  & $7.15\cdot10^{-4}$  & $6.11\cdot10^{-3}$  \\ \hline
			MOESP  & $9.11\cdot10^{-3}$ & $9.11\cdot10^{-3}$ &  $9.11\cdot10^{-3}$ &  $9.10\cdot10^{-3}$ &  $9.02\cdot10^{-3}$ &  $2.26\cdot10^{-2}$ \\ \hline
			CVA   &  $7.61\cdot10^{-3}$ &  $8.40\cdot10^{-3}$ &  $8.63\cdot10^{-3}$ & $8.95\cdot10^{-3}$ &  $9.03\cdot10^{-3}$ &  $8.76\cdot10^{-3}$	\\ \hline
		\end{tabular}
		\caption{Spiral inductor: A comparison of the $\cH_\infty$-norm of the error between the original and realized system using various approaches under different levels of noise in the measurement data.}
		\label{Table:HinfSpiral}
	\end{table}
\end{center}

\section{Discussion}
We would like to stress the point that we particularly focus on estimating the transfer functions at specific frequency points using the input and output data on a certain time interval with a specific sampling time. It is followed by employing the Loewner approach \cite{morMayA07} to construct a realization. 
The accuracy of the transfer function's estimation depends on the sampling time. Precisely, if the sampling time is small, we can estimate the transfer function more accurately. Furthermore, the number of samples also has an influence on the estimation, i.e., the more samples, the better the estimation of the transfer function. Last but not least, the design of the input should be such that it captures a wide range of frequencies or at least the range of frequencies of interest so that we have a good estimate of the transfer function in the desired range of the frequency. 

When applying the time-domain Loewner \cite{morBenGW15},
one has to consider that the input is carefully chosen to make the least-squares problem \eqref{LSprob2} solvable and the computation time relatively small.  The problem is set up carefully such that it allows us to estimate the transfer function at the specific frequency points that are also present in the input. Hence, the Loewner approach apparently is computationally cheaper as compared to MOESP and CVA methods. This is what we have observed in our numerical experiments as well. Moreover, we mention that MOESP and CVA approaches require a predefined order of the realization, and one typically determines a good order by trial and test; whereas, the Loewner offers an edge by allowing us to determine a suitable order by looking at the decay of the singular values of the Loewner pencil. Furthermore, we have studied the robustness of all three approaches, and our empirical study showed that these methods are quite robust to the various different levels of Gaussian white noise. 

\section{Conclusions}\label{sec:conclusion}
In this paper, we have discussed a procedure to inferring a port Hamiltonian realization using time-domain input-output data. For this, we have first estimated the frequency response data using time-domain input-output measurements. It is followed by employing the methodology proposed in \cite{morBenGV20} that yields a port-Hamiltonian realization. We have illustrated the proposed procedure using two numerical examples and compared with two other popular system identification techniques. In this study, we noticed the importance of the input choice. The input has to excite the frequencies that have the greatest impact on the system's response. If these are not known in advance, then one has to excite a large range of frequencies. As further research, one could further study the choice of the input and its composing frequency components in an adaptive way such that only the most contributing frequencies to the system response are considered.

\section*{Code availability}
A \matlab~implementation to reproduce the results presented in this paper can be found at \\ \url{https://gitlab.mpi-magdeburg.mpg.de/cherifi/ph-realizations-from-time-domain-data}.

\bibliographystyle{unsrt}
\bibliography{mor}

\end{document}